\documentclass[a4paper,onecolumn,12pt,unpublished]{quantumarticle}
\pdfoutput=1

\usepackage[utf8]{inputenc}
\usepackage[english]{babel}
\usepackage[T1]{fontenc}

\usepackage[numbers,sort&compress]{natbib}

\usepackage{hyperref}
\usepackage{tikz}
\usepackage{lipsum}

\usepackage{graphicx}
\usepackage{multirow}
\usepackage{amsmath}
\usepackage{amssymb}
\usepackage{amsfonts}
\usepackage{amsthm}
\usepackage{mathrsfs}
\usepackage{textcomp}
\usepackage{manyfoot}
\usepackage{booktabs}
\usepackage{algorithm}
\usepackage{algorithmicx}
\usepackage{algpseudocode}
\usepackage{listings}
\usepackage{braket}
\usepackage{array}
\usepackage{color,soul}
\usepackage{subcaption}

\begin{document}

%TO UPDATE - something like: Novel NISQ-friendly Coined Quantum Walk Algorithm for Chaos-based Key Obfuscation
\title{Novel Low-depth Coined Quantum Walk Algorithm for Symmetric-key Generation Applications}

\author{Natalie Gibson}
\orcid{0009-0008-8947-8065}
\email{natalie.gibson@aalto.fi}
\thanks{corresponding author.}

\author{Niklas Keckman}
\orcid{0009-0006-1865-2017}

\author{Andrea Marchesin}
\orcid{0000-0002-8194-7161}
\author{Matti Raasakka}
\orcid{0000-0003-2636-0429}
\author{Ilkka Tittonen}
\orcid{0000-0002-2985-9789}
\affiliation{Micro and Quantum Systems Group, Department of Electronics and Nanoengineering, Aalto University, Espoo 02150, Finland}

\begin{abstract}
    We present a novel lackadaisical alternating quantum walk (LAQW) algorithm whose circuit depth scales as $\mathcal{O}(n^2+nt)$ for a $2^n\times 2^n$-sized lattice over $t$ time steps. We show that this is a significant depth reduction compared to the existing controlled alternating quantum walk (CAQW) model, which has a circuit depth that scales as $\mathcal{O}(n^2t)$~\cite{li2017controlledalternatequantumwalks}. This makes the implementation of the LAQW viable for Noisy Intermediate-scale Quantum (NISQ) devices. We then showcase the applicability of the LAQW algorithm by proposing a chaos-based symmetric-key generation scheme. Our approach uses the LAQW as a quantum entropy source from which reproducible random bitstring sequences are generated using the underlying probability distribution and subsequent post-processing methods. We provide a comprehensive evaluation of the LAQW algorithm and demonstrate the reproducibility of 128-bit keys under simulated quantum noise provided by IBM's \texttt{FakeTorino} backend. A direct comparison with the CAQW model, which has been used in image obfuscation and hash function schemes~\cite{CAQWmed,AQWPRNG2020,li2017controlledalternatequantumwalks}, highlights the potential and usefulness of the LAQW model in cryptographic applications.
\end{abstract}

\section{Introduction}\label{sec1}
\label{introduction}

\textit{Discrete-time quantum walks} (DTQWs)~\cite{PortugalQWs,Venegas-Andraca2008} are quantum analogues of classical random walks, where a ``walker'' evolves on a graph under unitary coin and shift operators. This evolution generates interference-based probability distributions that are highly sensitive to initial conditions~\cite{Abd_El-Latif2020-hc} and can spread with exponential or quadratic speed-ups over the graph when compared to classical random walks~\cite{KADIAN2021100419}. These properties have led to DTQWs being used in multitude of application fields such as spatial search problems, optimization problems, graph theory and cryptography~\cite{KADIAN2021100419}. 

For cryptographic applications, DTQWs are a useful tool due to them being \textit{chaotic systems}~\cite{math11112585}. A chaotic system can be characterized by sensitivity to initial conditions, nonlinearity, and aperiodicity, all of which are properties that DTQWs exhibit. Chaotic systems can be used as \textit{symmetric-key generators}  for chaos-based symmetric obfuscation, which can be a cost-friendly alternative to traditional obfuscation in terms of computational power and computation time. Symmetric-key generators enable two communicating parties who share a small-sized secret key to independently generate an identical, large, shared key using a chaotic system. In the case of DTQWs, the initial parameters of the quantum walk are the smaller secret key, and the large shared key can be obtained using the probability distribution generated by the quantum walk.

Depending on the application field, there are many different variants of DTQW models that modify the coin space dimensionality, shift operator, or graph topology to achieve desired walk dynamics. A common choice for cryptographic applications is the \textit{alternating quantum walk} (AQW)~\cite{AQW2011}. The AQW reduces the dimensionality required for walks on two-dimensional lattices by alternating a walker's motion between two one-dimensional DTQWs on a cycle graph, where the graph nodes correspond to the walker's horizontal and vertical position on the lattice. A variant of the AQW is the \textit{controlled alternating quantum walk} (CAQW)~\cite{li2017controlledalternatequantumwalks}, which introduces an additional binary key string that regulates the coin operator at each time step of the walk. This increases the parameter space of the quantum walk, making the resulting probability distribution more unpredictable and sensitive to initial parameters. This, in turn, increases the applicability of the walk as a chaotic system.

In recent years, the chaotic dynamics exhibited by these models have motivated the use of DTQW-based image obfuscation schemes~\cite{CAQWmed,Abd_El-Latif2020-hc,Alblehai2025-fg,AQWPRNG2020,AQWAes,AQWQuantumIE,Yang2015-pf,AQWRubikscube}, hash functions~\cite{li2017controlledalternatequantumwalks}, and random number generators~\cite{AQWPRNG2020,Meng24,Sarkar2019-bc}. These applications primarily use AQWs or CAQWs as subroutines for symmetric obfuscation-deobfuscation schemes to produce large pseudorandom keys from a small set of initial values. These quantum walk-based schemes are reported to have strong statistical randomness properties and high sensitivity to initial conditions, which we will also discuss in detail in Section \ref{sec5}. However, a challenge for their practical implementation on near-term quantum processing units (QPUs) is to design efficient quantum circuits that have shallow depth and gate count. The larger the circuit depth and gate counts are, the more noisy the outcomes of the models are. If the outcomes are too noisy, the models cannot be used as a subroutine for obfuscation-deobfuscation schemes, since the results are not reproducible. 

 The quantum circuit implementation of these models requires a depth that scales as $\mathcal{O}(n^2 t)$ for a $2^n \times 2^n$-sized lattice over $t$ time steps, making it resource-intensive for noisy intermediate-scale quantum (NISQ) devices. Our work adapts the \textit{lackadaisical quantum walk} (LQW) model, originally proposed by~\cite{Wong_2015}, to study quantum walks on graphs with self-loops at each node of the graph. By incorporating self-loops at the nodes of the lattice, the \textit{lackadaisical alternating quantum walk} (LAQW) achieves a circuit depth that scales as $\mathcal{O}(n^2 + n t)$ for a $2^n \times 2^n$-sized lattice over $t$ time steps, while still being compatible for DTQW-based obfuscation schemes that would use the AQW or the CAQW models.  

In this work, we introduce and analyze a novel symmetric-key generation scheme, which serves as an application for our novel LAQW model. Specifically, we make the following contributions: first, we present the LAQW model and show that its quantum circuit implementation achieves a significant depth reduction compared to the CAQW used in ~\cite{CAQWmed,AQWPRNG2020,li2017controlledalternatequantumwalks}, increasing its viability for near-term QPUs. We then propose a reproducible scheme to generate bitstrings from the sampled LAQW probability distribution, incorporating a prime-modulus mapping to promote uniformity. Finally, we provide an evaluation of the scheme's cryptographic properties, including statistical randomness that can be checked through NIST tests~\cite{8966,233791}, key reproducibility, and resilience to quantum noise. 

This paper is organized as follows: in Section \ref{sec2}, we introduce the general coined DTQW model (Section \ref{sec2.1}). Then, the AQW and CAQW models are detailed (Section \ref{sec2.2}), followed by a discussion of their circuit implementations (Section \ref{sec2.3}). We then conclude the section by introducing the LQW model (Section \ref{sec2.4}). We start Section \ref{sec3} by proposing the novel LAQW model (Section \ref{sec3.1}) and present our optimized circuit construction of the LAQW model (Section \ref{sec3.2}). Section \ref{sec4} details the full proposed symmetric-key generation scheme, starting with the bitstring generation procedure from LAQW measurements (Section \ref{bitstring_generation}) followed by the bit extraction process (Section \ref{min_entropy}). In Section \ref{sec5}, we present the comprehensive results of our scheme. Section \ref{circuit_comparisons} compares the circuit depths between the LAQW and CAQW algorithms, followed by a discussion regarding the sensitivity of the quantum walk initial parameters and its importance for our chosen symmetric-key generation application. We then report on the statistical randomness evaluations of the generated bitstring sequences (Section \ref{statistical_analysis}), investigate reproducibility across trials (Section \ref{reproducibility}), and study noise resilience under realistic hardware conditions (Section \ref{effects_noise}). The paper concludes with an overview of considerations and future work in Section \ref{sec6}. 

\section{Background}\label{sec2}\label{background}

In this section, we review the pre-existing DTQW, AQW, CAQW, and LQW models and briefly discuss the quantum circuit implementation of these models. These models will be used as a basis for the novel lackadaisical alternating quantum walk (LAQW) model that will be proposed in Section \ref{sec3}.

\subsection{Discrete-time coined quantum walk model}\label{sec2.1}

The DTQW model is the quantum mechanical counterpart of the classical random walk model~\cite{PortugalQWs}. The DTQW can be conducted on several different graph topologies. In this paper, we consider the one-dimensional DTQW on a cycle graph of length $2^n$. Here, $n$ is the number of qubits that are used for representing the nodes of the graph. On this graph type, the DTQW can be represented as a series of unitary operations $U$ applied to a state vector $\ket{\psi}$ in the Hilbert space $\mathcal{H}=\mathcal{H}_p\otimes\mathcal{H}_c$. The position space $\mathcal{H}_p$ is spanned by $2^n$ basis vectors $\ket{i}$, where $i\in \{1,2,...,2^n\}$, that are associated with the possible positions of a quantum walker on the $N$ node cycle. The coin space $\mathcal{H}_c$ is spanned by two basis vectors $\ket{0}$ and $\ket{1}$ that determine whether the walker moves clockwise or counter-clockwise on the cycle graph respectively. The walker, starting at an initial state $\ket{\psi_0}\in \mathcal{H}$, takes $t$ steps when $U$ is applied to $\ket{\psi_0}$ $t$ times. The resulting state $\ket{\psi_t}=U^t\ket{\psi_0}$ can then be measured to obtain the position of the quantum walker after $t$ steps of the walk.

The $N$-node cycle graph can be expanded into an $N\times N$-sized two-dimensional lattice with periodic boundary conditions by considering two $N$ node cycle graphs that corresponds to the horizontal and vertical movement of the walker respectively \cite{PortugalQWs}. In the case of the two-dimensional periodic lattice, the position space $\mathcal{H}_p$ is spanned by basis vectors $\ket{x,y}$ where $x,y\in \{1,...,N\}$ that correspond to the position of the walker on the lattice. The coin space $\mathcal{H}_c$ is spanned by basis vectors $\ket{00}, \ket{01}, \ket{10}, \ket{11}$ that determine whether the walker moves up right left or down on the graph, respectively. Extending the graph structure into even higher dimensions follows the same principle.

\subsection{Alternating discrete-time quantum walk model}\label{sec2.2}

The AQW model was first proposed in~\cite{AQW2011} in order to reduce the dimension of the coin space that would be required to implement the DTQW on a two-dimensional periodic lattice. The AQW is implemented by alternating between two one-dimensional DTQWs on the cycle graph, with one cycle corresponding to the horizontal movement, and the other cycle corresponding to the vertical movement of the quantum walker. The same coin qubit is used for both walks, thus reducing the amount of required coin qubits to one. The unitary evolution operation $U$ in the AQW is defined as \begin{equation}
    U=S_v(I\otimes C_\theta) S_h(I\otimes C_\theta),\label{AQWModel}
\end{equation}
where $S_v$ and $S_h$ are shift operations corresponding to the vertical and horizontal movement of the quantum walker, respectively and $C_\theta$ is a coin operator that evolves the coin space. The shift operators are defined as \begin{equation}
    S_v=(\sum\limits_{y,x}\ket{x,y+1}\bra{x,y})\otimes\ket{0}\bra{0}+(\sum\limits_{y,x}\ket{x,y-1}\bra{x,y})\otimes\ket{1}\bra{1}
\end{equation} and \begin{equation}
    S_h=(\sum\limits_{y,x}\ket{x+1,y}\bra{x,y})\otimes\ket{0}\bra{0}+(\sum\limits_{y,x}\ket{x-1,y}\bra{x,y})\otimes\ket{1}\bra{1},
\end{equation} where $y$ and $x$ correspond to the vertical and horizontal position of the walker respectively. The coin operator  $C_\theta$ is defined as \begin{equation}
    C_\theta=\begin{bmatrix}
        \cos(\theta)&\sin(\theta)\\
        \sin(\theta)&-\cos(\theta)
    \end{bmatrix}\label{Cointheta},
\end{equation}
where the parameter $\theta$ determines how the coin qubit evolves. With different values of $\theta$, the coin operator changes the probability of the walker going either clockwise or counter-clockwise on the cycle graph~\cite{PortugalQWs}.

AQW models have been used in multiple different symmetric image obfuscation schemes~\cite{CAQWmed,AQWPRNG2020,AQWAes,AQWRubikscube} and hash function implementations~\cite{li2017controlledalternatequantumwalks}. Each of these approaches differ from one another drastically in terms of how the obfuscation-deobfuscation schemes are implemented. However, all of these approaches have in common that they use the AQW as a subroutine for producing random probability distributions on a two-dimensional lattice. Instead of using the AQW model as presented in Eq. \eqref{AQWModel}, many of these approaches consider the CAQW model, which is a slightly modified version of the AQW~\cite{CAQWmed,AQWPRNG2020,li2017controlledalternatequantumwalks}. In the CAQW, the unitary operation $U$ is defined as \begin{equation}
    U=S_v(I\otimes C_{\theta_i}) S_h(I\otimes C_{\theta_i}),\label{CAQWUnitary}
\end{equation}
where $C_{\theta_i}$ is determined by a bitstring of length $k=t$. First, two angles $\theta_0,\theta_1\in [0,2\pi]$ are chosen randomly from a uniform distribution on $[0,2\pi]$. Then, the corresponding coin operators $C_{\theta_0}$ and $C_{\theta_1}$ are determined as in Eq. \eqref{Cointheta}. For each step $j\leq t$ of the walk, the variable $i\in \{0,1\}$ has a value that corresponds to the $j$th bit of the bitstring. 

Whenever AQW or CAQW is used, the sizes of the cycle graphs corresponding to the horizontal and vertical components have to be odd~\cite{CAQWmed,AQWPRNG2020}. If even-sized cycle graphs are used, only half of the nodes in the graph will have a non-zero probability for the walker being at that node. To understand this behavior, assume that the walker on a $m$-node cycle starts in position $\ket{k}$, where $k\in\{0,1,...,m-1\}$, and $m$ is even. \begin{itemize}
    \item[$\cdot$] Assume that $k\in \{2,4,...,m-2\}$. Then, after one step of the walk the position of the walker is $\ket{k\pm 1}$, which is odd.
    \item[$\cdot$] Assume that $k=0$. Then, after one step of the walk the position of the walker is $\ket{m-1}$ or $\ket{1}$, which are both odd.
    \item[$\cdot$] Assume that $k\in \{1,3,...,m-3\}$. Then, after one step of the walk the position of the walker is $\ket{k\pm 1}$, which is even.
    \item[$\cdot$] Assume that $k=m-1$. Then, after one step of the walk the position of the walker is $\ket{m-2}$ or $\ket{0}$, which are both even.
\end{itemize}
 So, by starting from an even position, the walker always moves to an odd position. Similarly, by starting from an odd position, the walker always moves to an even position. Thus, at a certain time step, the walker is either completely on odd-numbered positions or completely on even-numbered positions. To see why this pattern does not emerge when $m$ is odd, consider the walker at position $\ket{0}$. After one step of the walk, the walker can be either at position $\ket{1}$ or position $\ket{m-1}$. Since $m$ was assumed to be odd, $\ket{m-1}$ is an even position. Therefore, with odd $m$, the walker can have a non-zero probability of being at even-numbered and odd-numbered nodes at the same time step.

Probability distributions where only half of the nodes are non-zero should be avoided. Otherwise, the probability distribution is more predictable, since the probabilities of half of the nodes are known. Additionally, some properties of the initial parameters of the walk, such as whether $t$ is even or odd, can be deduced from the probability distribution in this case. When AQW and CAQW are used for cryptographic purposes, both of these observations that occur when $N$ is even may compromise the security of the cryptographic methods. This is why an odd-sized cycle graph is required when using AQW and CAQW.

\subsection{Circuit implementation of discrete-time coined quantum walks}\label{sec2.3}

Different circuit implementations of DTQWs have been studied in order to perform DTQWs on NISQ-era QPUs. Multi-controlled X-gate (MCX) based shift operators were first proposed in~\cite{douglas2009efficientquantumcircuitimplementation}. An alternative version of the shift operators is presented in~\cite{Georgopoulos_2021}, which uses rotation gates instead of MCX-gates in order to reduce the amount of ancilla qubits needed. A Quantum Fourier transform (QFT)-based approach is given in~\cite{koch2021gatebasedcircuitdesignsquantum} and is further improved upon in~\cite{Razzoli_2024}. This improved approach uses the QFT in order to reduce the circuit depth required for implementing the increment and decrement shift operators. 

The increment and decrement shift operations on a periodic line of size $n$ can be related to the shift operator $S$ by \begin{equation}
    S=\begin{bmatrix}
        P_0&0\\0&P_1
    \end{bmatrix},
\end{equation} where $0$ is the $n\times n$ null matrix and the matrices $P_0$ and $P_1$ are the decrement and increment matrices, respectively. They are defined as 

\begin{equation}
    P_0=\begin{bmatrix}
        0&1&0&\dots&0\\
        0&\ddots&\ddots&\ddots&\vdots\\
        0&\ddots&\ddots&1&0\\
        \vdots&\ddots&0&0&1\\
        1&\dots&0&0&0
    \end{bmatrix}, \quad P_1=\begin{bmatrix}
        0&0&0&\dots&1\\
        1&\ddots&\ddots&\ddots&\vdots\\
        0&\ddots&\ddots&0&0\\
        \vdots&\ddots&1&0&0\\
        0&\dots&0&1&0
    \end{bmatrix}.
\end{equation}
Implementing these matrices with MCX gates can be done using a quantum circuit as shown in~\cite{Georgopoulos_2021} with a circuit depth of $\mathcal{O}(n^2)$. Thus, by considering $t$ discrete time steps, the depth of the DTQW circuit when using this approach will be $\mathcal{O}(n^2t)$.

The QFT-based approach of~\cite{koch2021gatebasedcircuitdesignsquantum} and~\cite{Razzoli_2024} has a circuit depth $\mathcal{O}(n^2+nt)$. This circuit depth is an improvement when compared to the MCX gate-based approach of~\cite{Georgopoulos_2021} for large $t$. However, the QFT-based approach requires the increment and decrement matrices $P_0$ and $P_1$ to be circulant. This is due to circulant matrices being diagonalized by the matrix $\mathcal{F}$ corresponding to the QFT as \begin{equation}
    P_0=\mathcal{F}^\dagger\Omega^\dagger\mathcal{F},\;P_1=\mathcal{F}\Omega \mathcal{F}^\dagger, \label{PMatDiag}
\end{equation} where $\Omega=\text{diag}(1,\exp(2\pi i/2^n)^1,...,\exp(2\pi i/2^n)^{2^n-1})$. This way, whenever $P_0$ or $P_1$ is applied $t$ times in a row, we can instead apply \begin{equation}
P_0^t=\mathcal{F}^\dagger(\Omega^\dagger)^t\mathcal{F},\;P_1^t=\mathcal{F}\Omega^t \mathcal{F}^\dagger.
\end{equation} 
If $P_0$ and $P_1$ are not diagonalized by $\mathcal{F}$, this approach does not work. Therefore, the increment and decrement matrices are required to be circulant.

When considering the implementation of odd-sized cycle graphs using qubit systems, the matrices $P_0$ and $P_1$ cannot be circulant. To see why this is true, note first that the matrices have to be even-sized when a composite system of qubits is considered, since the dimensionality of the Hilbert space is $2^n$, where $n$ is the number of qubits. In order to implement an odd-sized cycle, an odd number of rows of $P_0$ and $P_1$ would have to be non-zero. This would mean that at least one of the rows of $P_0$ or $P_1$ would have to be entirely empty. However, for $P_0$ and $P_1$ to be circulant, each row of $P_0$ and $P_1$ would then need to be empty. Thus, $P_0$ and $P_1$ cannot be circulant if the cycle graph is odd-sized. 

This means that the QFT-based approach cannot be used on odd-sized cycle graphs, and therefore, AQW and CAQW circuits cannot be constructed using the QFT-based approach of~\cite{koch2021gatebasedcircuitdesignsquantum} and~\cite{Razzoli_2024}. 

\subsection{Lackadaisical discrete-time quantum walk}\label{sec2.4}

Lackadaisical discrete-time quantum walks (LQWs) are DTQWs where an additional self-loop is introduced to every node of the graph upon which the walk takes place. This allows the walker to stay in place with some probability, which makes the dynamics of the walk substantially different from regular DTQWs. Mostly, LQWs have been considered for spatial search problems. The application of LQWs to spatial search was first introduced in~\cite{Wong_2015}. Afterwards, applications of LQWs for spatial search of multiple marked states that are located next to each other have been studied in~\cite{saha2018searchclusteredmarkedstates} and applications of LQWs for spatial search of multiple arbitrarily placed marked states have been investigated in~\cite{LAQWmultipleTargs}. Little to no research has been conducted in using LQWs for obfuscation applications.

\section{Methods}
\label{sec3}

In this section, as the main contribution of this paper, we propose a novel lackadaisical controlled alternating quantum walk (LAQW) model that can be used for the same application purposes as the AQW and CAQW models, while having a significantly shallower circuit implementation. 
We then show the quantum circuit implementation of this LAQW model.

\subsection{Lackadaisical alternating quantum walk}\label{sec3.1}

In order to use even-sized cycle graphs while still producing non-predictable patterns with the DTQW, we propose a modification to the traditional CAQW by introducing a self-loop to each of the nodes of the cycle. We call this the \textit{lackadaisical alternating quantum walk} (LAQW). For each time step of the LAQW, the walker on the cycle graph has the option to move left, right, or stay in place. With this modification to the CAQW, the walker starting at any node of the cycle graph of size $2^n$ can have a non-zero probability of being at any node of the graph after $t> 2^n$ steps of the walk. As explained previously, the walker only has a non-zero probability of being at half of the nodes of the same cycle graph at time $t$ for the AQW and CAQW models. 

By adding a second coin qubit, the shift operator $S_h$ can be defined as 
\begin{align}
S_h =(\sum\limits_{y,x}\ket{x+1,y}\bra{x,y})&\otimes\ket{00}\bra{00})+(\sum\limits_{y,x}\ket{x-1,y}\bra{x,y})\otimes\ket{11}\bra{11}\notag \\
&+I\otimes(\ket{01}\bra{01}+\ket{10}\bra{10}).
\label{LAQWShift}
\end{align}
 From Eq. \eqref{LAQWShift}, it can be seen that the position of the walker will only move if the state of the coin qubits is $\ket{00}$ or $\ket{11}$. Otherwise, the walker will stay in place. The shift operator $S_v$ is defined in a similar way. By construction, the shift matrices $P_0$ and $P_1$ will be circulant. This way, we are able to use the QFT-based approach of~\cite{Razzoli_2024}, while still producing non-periodic patterns at the cost of adding a single qubit to the coin space. Using this approach, the LAQW has lower circuit depth than the CAQW. The circuit depths of the LAQW and CAQW will be analyzed and compared in Section \ref{circuit_comparisons}.

\subsection{Circuit implementation of the LAQW}\label{sec3.2}

The proposed LAQW algorithm works as follows: the walk is first initialized according to the initial parameters of the walk. Then, the walk is evolved for $t$ time steps, after which the position of the walker can be measured. A single iteration of this procedure will be referred to as a \textit{shot}. The algorithm will be executed for a large number of shots, after which the frequencies of measurements at different nodes will give an approximation of the underlying probability distribution that the LAQW produces. This procedure of obtaining this probability distribution will be referred to as a \textit{run}. 

\begin{figure}[t]
    \centering
    \includegraphics[width=\linewidth]{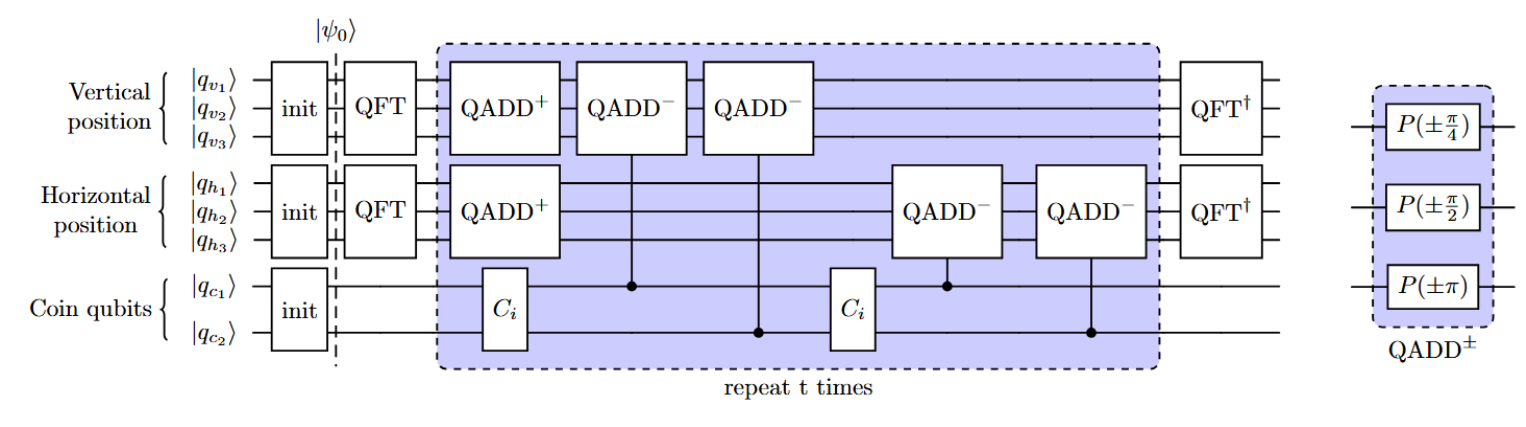}
   \caption{Circuit implementation of the LAQW on an $8\times 8$-sized two-dimensional lattice, and the circuit implementation of the $\text{QADD}^\pm$ gates for cycle graphs of $8$ nodes.
    }
    \label{fig:LAQWCircuit}
\end{figure}

A circuit implementation of the proposed LAQW algorithm can be seen in Figure \ref{fig:LAQWCircuit}. The gates labeled ``init'' correspond to the gates required to create the initial state of the walk, $\ket{\psi_0}$. For the position qubits, this is done by applying $X$-gates to the appropriate qubits. For the coin qubits, the initial state, as detailed in Table \ref{tab:Parameters}, can be created by applying an $R_y$-gate with an angle $2\alpha$ to the qubits. The gates labeled $C_i$ correspond to the two coin operators $C_0$ and $C_1$, as described in Table \ref{tab:Parameters}. The part of the circuit inside the highlighted box will be repeated $t$ times, corresponding to $t$ time steps of the LAQW. The gates labeled $\mathrm{QADD}^+$ and $\mathrm{QADD}^-$ are used to implement the matrices $\Omega$ and $\Omega^\dagger$ from Eq. \eqref{PMatDiag}, respectively. They can be constructed by applying phase gates \begin{equation}
    P(\theta)=\begin{bmatrix}
    1&0\\0&\exp(i\theta)
\end{bmatrix}
\end{equation} 
with different angles $\theta$, as described in ~\cite{koch2021gatebasedcircuitdesignsquantum}. It is important to note that the $\mathrm{QADD}^+$ and $\mathrm{QADD}^-$ gates are implemented by applying a single phase gate to each of the qubits, and thus, the gates can be constructed in constant circuit depth. Single qubit-controlled versions of these gates have $\mathcal{O}(n)$ circuit depth. 

\begin{table}[h]
\centering
\begin{tabular}{|c|l|}
\hline
\textbf{Parameter} & \textbf{Description}  \\ \hline
$x_0$  & \begin{tabular}[c]{@{}l@{}}
Determines the initial position of the walker on the \\ cycle graph corresponding to the horizontal movement.
\end{tabular} \\ \hline
$y_0$  & \begin{tabular}[c]{@{}l@{}}
Determines the initial position of the walker on the \\ cycle graph corresponding to the vertical movement.
\end{tabular} \\ \hline
$\alpha$  & \begin{tabular}[c]{@{}l@{}}
Determines the initial state of the coin qubits such that\\ $\ket{q_{c_1}}=\ket{q_{c_2}}=\cos(\alpha)\ket{0}+\sin(\alpha)\ket{1}$
\end{tabular} \\ \hline
$\theta_0$ & \begin{tabular}[c]{@{}l@{}}
Determines a coin operator $C_0$ according to Eq. \ref{Cointheta}
\end{tabular} \\ \hline
$\theta_1$ & \begin{tabular}[c]{@{}l@{}}
Determines a coin operator $C_1$ according to Eq. \ref{Cointheta}
\end{tabular} \\ \hline
$t$ & \begin{tabular}[c]{@{}l@{}}
Determines the number of steps the walk will take.
\end{tabular} \\ \hline
$\mathcal{K}$ & \begin{tabular}[c]{@{}l@{}}
A bitstring of length $t$. The value $j\in \{0,1\}$ of the bitstring \\at position $i$ determines the coin operator $C_j$ that will be\\ used on step $i$ of the walk.
\end{tabular} \\ \hline
\end{tabular}
\caption{Parameters of the LAQW algorithm.}
\label{tab:Parameters}
\end{table}

Our proposed LAQW model has seven parameters that each affect the final probability distribution produced by the walk. These parameters and their descriptions can be seen in Table \ref{tab:Parameters}.

\begin{figure}[t]
    \centering
    \includegraphics[width=\linewidth]{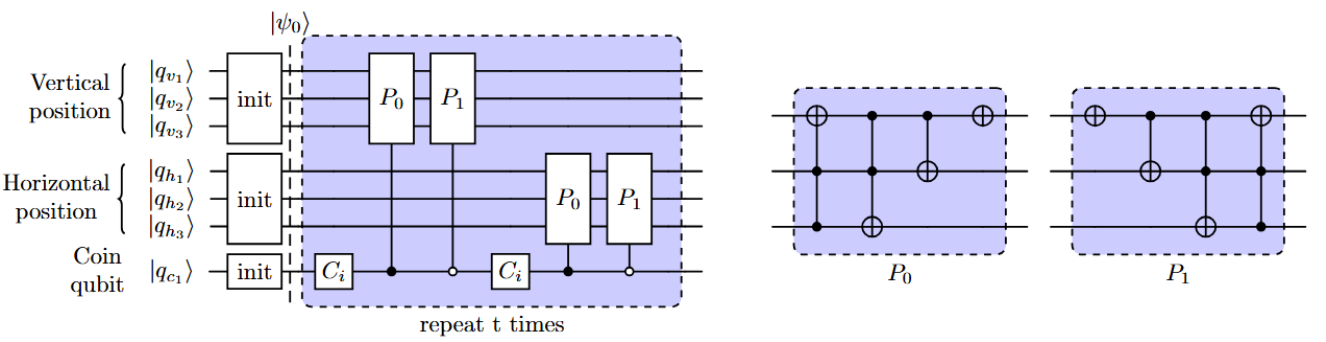}
    \caption{Circuit implementation of the CAQW circuit on a $7\times 7$-sized lattice, and the circuit implementation of the $P_0$ and $P_1$ matrices for a $7$-node cycle graph.}
    \label{fig:IncDecCirc}
\end{figure}

Figure \ref{fig:IncDecCirc} shows the implementation of the CAQW circuit for a $7\times 7$-sized lattice. In the same figure, the implementation of the increment and decrement matrices, $P_0$ and $P_1$, for this odd-sized lattice can be seen. The gates labeled ``init'' and the gates labeled $C_i$ are constructed in the same way as in the LAQW circuit. The CAQW circuit will be used as a comparison when showcasing the circuit depth of the LAQW circuit and the sensitivity of the LAQW model to the initial walk parameters.

\section{Symmetric-key generation application}\label{sec4}

This section details our symmetric-key generation methodology using the LAQW. The process consists of two main stages: first, a ``raw'' bitstring is constructed from the LAQW probability distribution using post-processing methods described in Section \ref{bitstring_generation}. Then, we estimate the conservative \textit{minimum entropy} of the raw bitstring and extract a smaller-sized key of chosen length by performing the a weighted bit extraction scheme, which will be described in detail.

Entropy quantifies the fundamental uncertainty in a random process~\cite{233791}. In the context of cryptography, a high-entropy source corresponds to the difficulty an adversary faces in guessing its output, making it a fundamental topic for secure key generation. Although there are several entropy measures, \textit{min-entropy}~\cite{9393401} provides the most conservative security estimate by bounding the probability of guessing the most likely outcome. If the min-entropy of a source is too low, an adversary could determine its output with non-negligible probability, making the source insecure to use for symmetric-key generation. The following subsections elaborate on each step of the scheme.

\subsection{Bitstring generation}
\label{bitstring_generation}

The raw bitstring is obtained through multiple post-processing methods applied to the probability distribution. Starting from the measured counts of each lattice position, we round each count to integer intervals and then map those integers to byte values via a prime-modulus mapping, designed to ensure uniformity in values. Each step of the bitstring generation is now described in detail. 

\paragraph{Overview of the scheme}

\begin{figure}[t!]
    \centering
    \begin{subfigure}[t]{0.48\linewidth}
        \centering
        \includegraphics[width=\linewidth]{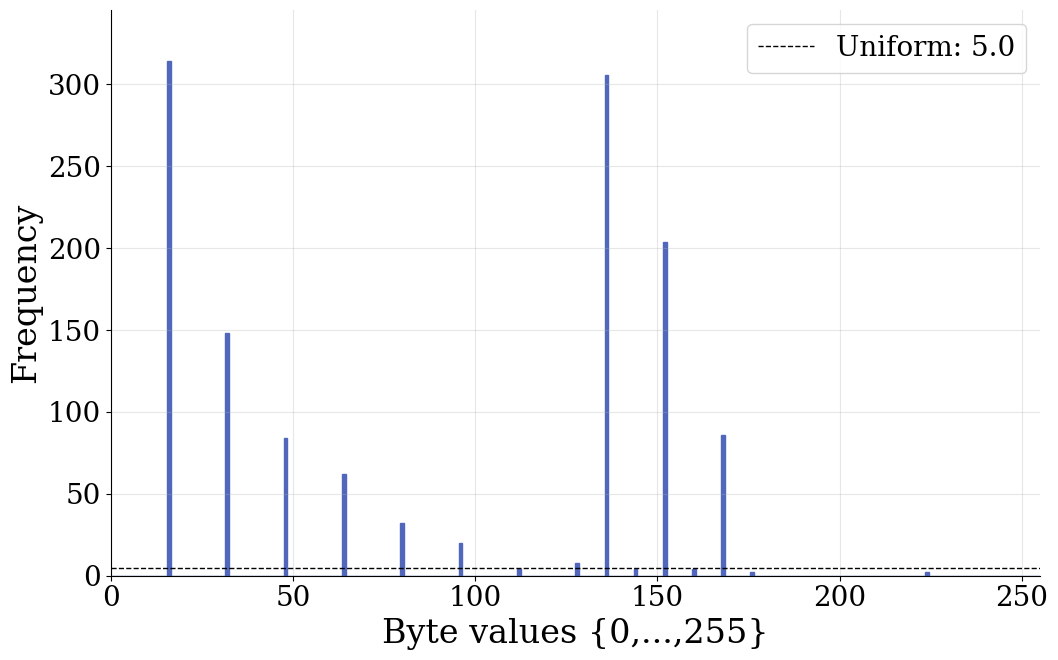}
        \caption{}
        \label{fig:256Mapping}
    \end{subfigure}
    \hfill
    \begin{subfigure}[t]{0.48\linewidth}
        \centering
        \includegraphics[width=\linewidth]{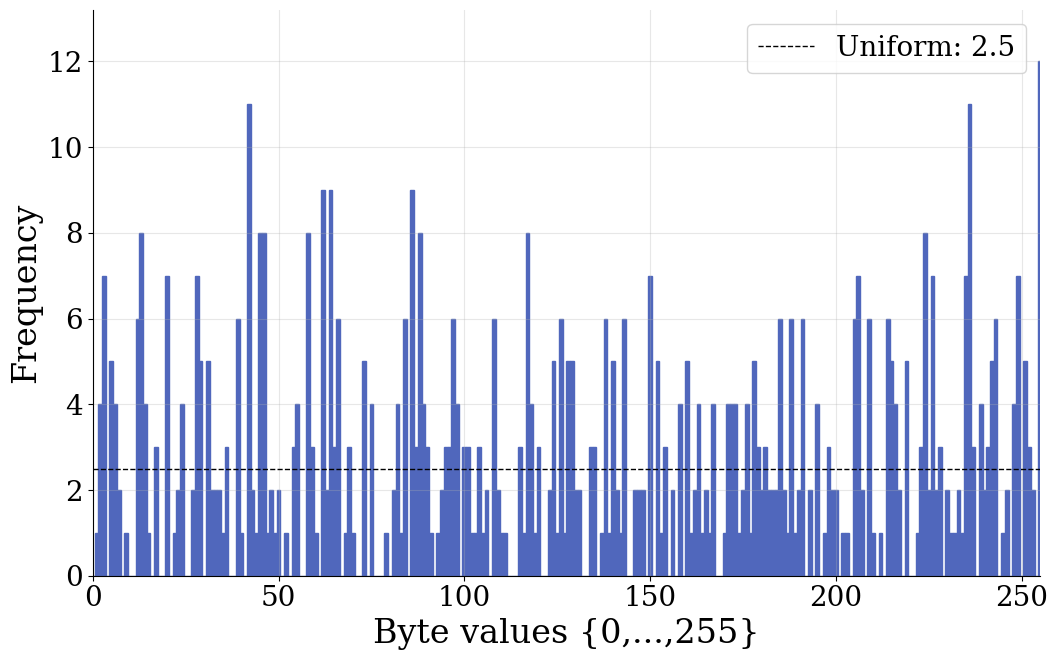}
        \caption{}
        \label{fig:primeMapping}
    \end{subfigure}

\caption{Resulting byte mapping distribution of the modulus 256 mapping (a) and the prime-modulus mapping (b). The total variation distance (TVD) from uniform of (a) is $0.948$ and the total fraction of ones is $0.227$. For (b), the TVD is $0.405$ and the total ones fraction is $0.511$.}
\end{figure}

\begin{enumerate}
    \item \textbf{Choose initial parameters:} We fix the initial walker position as a pair of randomly chosen horizontal and vertical coordinates $(x,y) \in \{1,\dots,2^{n}\}$ on a $2^n \times 2^n$ lattice. Floating-point values for the coin rotation parameters $\alpha, \theta_0, \theta_1 \in (0, \pi)$ are also chosen randomly, together with the number of time steps $t$. $t$ is chosen to be greater than $2^n$ to ensure the possibility of the walker to explore every node on the lattice. However, in order to minimize circuit depth, $t$ should be as small as possible. The binary key string parameter $\mathcal{K}$ should be chosen such that the number of bits is the length of $t$.
    
    \item \textbf{Collect measurements:} For each run index $r \in \{1,\dots,R\}$ and lattice position $i\in \{1,\dots,2^{n}\}$, we collect the \textit{counts} $c_{r,i}\in\{0, 1, \dots, S\}$ at each lattice node after making $S\approx10^6$ measurements. We define counts as the number of times the walker is observed at lattice position $i$ in run $r$. 
    
    \item \textbf{Coarse interval rounding:} We implement a deterministic interval rounding technique that allows for measured counts to be reproducible across repeated experiments when executing the quantum walk with identical initial parameters. If interval rounding is not applied, then it is unlikely to obtain the same probability distribution from the sampled measurements. Although the probability distribution remains stable for large shot counts $S$, individual lattice node counts can contain minor statistical fluctuations. Therefore, we map the counts to many intervals derived from the total shot count. The rounding procedure operates as follows: for each count $c_{r,i}$, we compute the rounded value as 
    \begin{equation}
        \widetilde{c}_{r,i} =
        \begin{cases}
            \left\lceil \dfrac{c_{r,i}}{\alpha S} \right\rceil \alpha S & \text{if} \hspace{2mm} c_{r,i} \leq 0.03 S, \\
            \left\lfloor \dfrac{c_{r,i}}{\beta S} + 0.5 \right\rfloor \beta S & \text{otherwise},
        \end{cases}
    \end{equation}
    where $\alpha = 0.005$ and $\beta = 0.010$ define the interval sizes as fractions of the total shot count $S$. This approach provides finer granularity for low-probability positions $c_{r,i} \leq 0.030S$, while implementing coarser intervals for higher-count nodes.
    
    \item \textbf{Convert integer counts to 8-bit ``blocks'':} The next step is to map the rounded counts $\widetilde c_{r,i}$ to  bytes $b_{r,j} \in \{0,\dots,255\}$. We first apply a deterministic mapping, where a specific prime modulus for each position is derived from the probability distribution. This approach was chosen since the rounded counts are inherently even-valued. The specific implementation of the prime-modulus mapping is as follows:
    \begin{enumerate}
        \item[(i)] For run $r$, the set of positions is arranged by the rounded counts $\widetilde{c}_{r,i}$ in ascending order into a list $\mathcal{I}_r$. This creates a permutation of the indices $i$ that are based on the probability distribution of that run.
        \item[(ii)] A fixed ordered list of prime numbers $m_i$ is predefined as $\mathcal{P} = (m_1,m_2,\dots, m_{2^n})$, where $m_1$ is the smallest prime and $m_{k+1}$ is the next consecutive prime after $m_k$ for $k=\{1,\dots,2^n-1\}$. The larger the value of $m_1$, the larger the amount of values that each node can obtain. Then again, the primes $m_k$ should be smaller than $256$ so that the values of the nodes can be represented as 8-bit integers. If the size of the list is such that the prime values surpass $256$, $m_k$ should be set again to $m_1$, with the following values repeating $m_2,\dots,m_{2^n}$  until the size of the list is $2^n$.
        \item[(iii)] The sorted list of $\mathcal{I}_r$ is used to index into the prime list $\mathcal{P}$. The $k$-th smallest count in sorted $\mathcal{I}_r$ is paired with the prime value $m_k = \mathcal{P}[k]$.
        \item[(iv)] The byte value for the $j$-th position, which occupies the $k$-th position in the sorted order, is then computed as
    \begin{equation}
        b_{r,i}=\bigg\lfloor\frac{(\widetilde{c}_{r,i}\mod{m_k})}{m_k - 1} \cdot 255 + 0.5 \bigg\rfloor \in \{0,\dots,255\}.
    \end{equation}
    \end{enumerate}
    By using a simple modulo 256 operation, where the rounded counts from step 3. are scaled by $S+1$ and then reduced modulo 256 to get the remainder, the resulting byte distribution would be highly non-uniform, as can be seen in Figure \ref{fig:256Mapping}. However, Figure \ref{fig:primeMapping} shows how the use of prime moduli disrupts this underlying structure and creates a near-uniform distribution of output bytes. The list $\mathcal{P}$ is a deterministic function of the underlying probability distribution and is non-injective. Therefore, the probability distribution cannot be retrieved based on the key string. 
    Each byte $b_{r,i}$ is then converted to an $8$-bit bitstring block, $\text{bin}_8(b_{r,i})$.
    
    \item \textbf{Concatenate to form final bitstring:} For each run $r$, we concatenate the produced bitstring blocks over all nodes to form a \textit{per-run} bitstring, $B_r$. The final ``raw'' bitstring is then 
    \begin{equation}
        B_{\text{raw}} = \underset{r=1}{\overset{R}{\parallel}}B_r = B_1 \parallel B_2 \parallel \dots \parallel B_R.
    \end{equation}
\end{enumerate}

\subsection{Min-entropy estimates and bit extraction}\label{min_entropy}

While the post-processing steps after the sampling of the quantum walk distribution deterministically transform the raw counts to produce a more uniform byte distribution, they do not create new entropy~\cite{Tomamichel_2011}. Any apparent increase in min-entropy calculated from the final bitstring would only be a reflection of the quality of the post-processing, not an indication that the quantum source has created additional randomness. Therefore, to establish a conservative baseline of the min-entropy of the LAQW, we evaluate the min-entropy on the probability distribution counts $c_{r,i}$ from a single LAQW run. This captures the randomness inherent to the quantum walk process prior to deterministic transformations.

\paragraph{Min-entropy estimation}
The maximum probability $p_{\max}$ of any output symbol is inferred from the observed counts of a quantum walk run. Given that $c_{\max}$ is the maximum count observed across the run, and $S$ is the total sample size, the empirical estimate is $\hat{p}_{\max} = c_{\max} / S$. Then, to establish a conservative upper bound on $\hat{p}_{\max}$, we construct a $(1 - \alpha)$ confidence interval for the binomial parameter which corresponds to the most frequent outcome. Specifically, we are interested in $p_{\max}^{\mathrm{ub}}$ that satisfies 
\begin{equation}
	\mathrm{Pr}(\hat{p}_{\max} \leq p_{\max}^{\mathrm{ub}}) \geq 1 - \alpha.
\end{equation}
While there are several methods for constructing binomial confidence intervals, we use the Clopper--Pearson method~\cite{be7c0fd0-f562-39ad-b8e0-716a276561d1, Darscheid2018-sj}. Unlike methods which give an approximation to the binomial distribution, the Clopper--Pearson provides the exact confidence interval obtained by inverting the cumulative binomial distribution. The upper bound corresponds to the upper endpoint of the Clopper--Pearson interval, given by
\begin{equation}
	p_{\max}^{\mathrm{ub}} = B_{1-\alpha} (c_{\max} + 1, S - c_{\max}),
\end{equation}
where $B_\gamma(\alpha, \beta)$ is the $\gamma$-quantile of the Beta distribution. The parameters $c_{\max} + 1$ and $S - c_{\max}$ correspond to the observed successes and failures, respectively, arising from the inversion of the binomial cumulative distribution function. By using this method, $\hat{p}_{\max}$ may be overestimated and the resulting min-entropy estimate will be conservative. However, this is desirable in our case, since a lower estimate is more secure. The approach of the NIST SP 800-90B publication~\cite{233791} uses a normal approximation to estimate the upper confidence bound, where 
\begin{equation}
	p_{\max}^{\mathrm{ub}} \approx \hat{p}_{\max} + z\sqrt{\frac{\hat{p}_{\max}(1 - \hat{p}_{\max})}{S - 1}}
\end{equation}
and $z=2.576$. For larger sample sizes, this estimate is closer to the value obtained with the Clopper--Pearson method, but $p_{\max}^{\mathrm{ub}}$ may be slightly underestimated, leading to a less conservative min-entropy estimate. \\

The min-entropy for each run is then found by 
\begin{equation}
    h_{\min} = -\log_2\big(p_{\max}^{\mathrm{ub}}\big),
\label{eq:hmin}
\end{equation}
with the final conservative min-entropy being the minimum value of $h_{\min}$ obtained across all runs. An upper bound $K_{\mathrm{raw}}$ is now established for the number of bits from the raw bitstring $B_{\text{raw}}$ that may be considered per run, 
\begin{equation}
\label{eq:Kraw}
    K_{\mathrm{raw}} = h_{\min}\cdot 2^n.
\end{equation}
The total number of bits across $r$ runs is then $K_{\mathrm{raw}} \cdot r$. From this Eq. \eqref{eq:Kraw}, along with our proposed bit extraction scheme, we can generate a random key that could be used in various other obfuscation schemes. \\

It is common in bit extraction schemes to use the \textit{Leftover Hash Lemma} (LHL)~\cite{Barak_2011, Tomamichel_2011} to determine an information-theoretic upper bound on the number of nearly-uniform bits $m$ that can be extracted from a source with min-entropy $h_{\mathrm{min}}$. For a chosen security parameter $\varepsilon > 0$ that bounds the statistical distance from uniformity, a universal hash function can be applied to $K_{\mathrm{raw}}$ to extract a number of bits, with the maximum number defined as 
\begin{equation}
    m_{\max} \le K_{\mathrm{raw}} - 2 \log_2\Big(\frac{1}{\varepsilon}\Big).
\end{equation}
The second term is the entropy loss required to ensure that $m_{\max}$ bits are $\varepsilon$-close to uniform. \\

However, in our case, we do not use a universal hash function. This is in part due to the fact that they require using a random, secret seed that grows with the number of extracted bits and which cannot be reused across different extractions. This adds a new, large secret that must be shared in order to retain key reproducibility, which unnecessarily complicates our goal of creating a reproducible mapping from initial quantum walk parameters to a final key. Therefore, in the following, we implement a deterministic \textit{weighted bit extraction} scheme that selects bits from the raw bitstring in a parameter-dependent way to construct a final key of length $m \le m_{\max}$. However, it should be made clear that our choice to not use a universal hash function means that the LHL does not hold in our case. This means that the uniformity of the resulting key cannot be inferred by the min-entropy estimate. Instead, the min-entropy should be viewed as a way to limit the number of bits to be extracted, but not as a guarantee on the quality of those bits. Establishing the latter for our extraction method would require a separate analysis that is beyond the scope of this work.

\paragraph{Weighted bit extraction}
We propose and implement a deterministic extraction scheme that allocates bits from each of the $r$ independent LAQW runs. The allocation is weighted by the initial coin state parameter $\alpha_r$ of each run. From numerical experiments, using this parameter as the weighting factor results in the raw and key bitstrings being the most \textit{reproducible}. Details on how the reproducibility is measured is described in the Results section (Section \ref{reproducibility}). For run $r$, we define an integer weight
\begin{equation}
    w_r = \lfloor S \alpha_r \rfloor
\end{equation}
and compute the total weight 
\begin{equation}
    W = \sum_{r=1}^R w_r .
\end{equation}
By including a weighting factor into the scheme that is based on an initial parameter value, the number of bits from each run that contribute to the final key will differ if $\alpha_r$ is changed by a small amount. Specifically, the number of bits $m_r$ extracted from the bitstring $B_r$ of run $r$ is decided by
\begin{equation}
    m_r = 
    \begin{cases}
        \left\lfloor \dfrac{w_r m}{W} \right\rfloor, & r = 1, \dots, R-1 \\[8pt]
        m - \sum\limits_{q=1}^{R-1} m_q, & r = R
    \end{cases}
\end{equation}
ensuring $\sum_{r=1}^R m_r = m$. The final key is then the concatenation of the first $m_r$ bits from each per-run bitstring $B_r$. This scheme is fully reproducible so long as identical initial parameters are used.

\section{Results and Comparisons}
\label{sec5}

This section first discusses the circuit comparisons between the LAQW and CAQW models and their viability on NISQ devices. We then compare the initial parameter sensitivity of the proposed LAQW model and the CAQW model of ~\cite{CAQWmed,AQWPRNG2020}. This is an important property to study, since a greater average value in the difference between slightly-modified probability distributions implies that there are a greater number of unique probability distributions that the model is capable of producing. In the case of symmetric-key generation, having high parameter sensitivity provides evidence that there is a large number of possible keys, making it harder for one to predict. Furthermore, comparing parameter sensitivity also allows us to better understand the dynamical behavior of the different quantum walk models.

In Section \ref{statistical_analysis}, we present an empirical evaluation of our LAQW-based symmetric-key generation scheme by first discussing the classical randomness properties of the generated raw and key bitstrings using tests from~\cite{8966}. We then discuss the reproducibility of the bitstrings in Section \ref{reproducibility}, as it is another important requirement for symmetric-key generation schemes. Furthermore, we investigate how the addition of realistic quantum hardware noise affects key reproducibility by using the \texttt{FakeTorino} backend, which mimics the noise and topology present in the real IBM Torino backend~\cite{IBMQuantum}. 

At the end of Sections \ref{statistical_analysis} and \ref{reproducibility}, we compare the statistical results of our proposed LAQW key generation scheme with an identical key generation scheme using the CAQW model discussed in Section \ref{sec2.2}. Both models use the same initial parameters ($\alpha, \theta_1, \theta_2, \mathcal{K}, t$) and post-processing methods detailed in Section \ref{bitstring_generation}, such as interval rounding, prime-modulus mapping, and weighted key extraction. Some of the main differences between the two quantum walk models are the size of the lattices on which the LAQW and CAQW run ($8\times8$ and $7\times7$, respectively) and the shift operators used in each walk, as described in Section \ref{circuit_comparisons}. The difference in lattice sizes also results in the length of the CAQW raw bitstring being smaller than for the LAQW raw bitstring. 

By comparing more than one quantum walk model, we aim to explore the potential benefits and trade-offs of incorporating different quantum walk models in cryptographic applications, such as symmetric-key generation or pseudorandom number generation.  

\subsection{Circuit Comparisons}
\label{circuit_comparisons}

Current NISQ-era quantum computing devices have many limitations in their capabilities. The main limitations consist of the number of qubits in the device, the qubit connectivity of the device, qubit decoherence times and quantum gate operation errors ~\cite{Corcoles_2020}. The effect of the circuit depth is directly related to the qubit decoherence as deeper circuits require more time to execute making the excited qubit state collapse into a basis state more likely. Circuit depth is also directly related to the problem of quantum gate operation errors, as deeper circuits have generally more quantum gates than shallow circuits.

The current quantum devices of IBM and Google are capable of handling circuit depths of over 100 gates only for a couple of qubits. As a comparison, a hypothetical 100 qubit quantum computer with 0.01\% error rate and perfect qubit readout would be capable of handling circuit depths of around 1 000 gates for up to 10 qubits ~\cite{proctor2024benchmarkingquantumcomputers}. For an $8\times 8$-sized and a $7\times 7$-sized lattices, the number of qubits required to implement the LAQW and CAQW algorithms respectively are 8 and 7, as can be seen from Figures \ref{fig:LAQWCircuit} and \ref{fig:IncDecCirc}. Since the required qubit amounts are low, the limiting factor is the circuit depth required to implement these algorithms. This is what the proposed LAQW model attempts to address.

When compared to the CAQW model, a key benefit of our model lies in its depth efficiency, making it a more practical candidate for implementation on current-generation quantum processors that are constrained by noise and limited circuit depth. In this section, we analyze and compare the circuit depth and total gate count of the LAQW (Figure \ref{fig:LAQWCircuit}) and CAQW (Figure \ref{fig:IncDecCirc}) model of~\cite{li2017controlledalternatequantumwalks}. The lattice sizes were kept relatively small due to the CAQW model's circuit depth scaling quadratically as $\mathcal{O}(n^2t)$. The circuits were then transpiled for a noiseless \texttt{GenericBackendV2} simulator and \texttt{FakeTorino}, which simulates IBM's \texttt{Torino} QPU~\cite{IBMQuantum}. This was done to investigate how lower connectivity of qubits affects the circuit depths. The \texttt{GenericBackendV2} simulator has an all-to-all qubit connectivity, whereas IBM's \texttt{Torino} QPU has heavy-hex topology, where each qubit is connected to three other qubits at maximum. For both models, we compiled $10$ separate circuits, each corresponding to a different number of time steps $t \in \{8, \dots, 20\}$. Since the number of shots needed for each run to obtain a stable probability distribution with our setup is $S \geq 10^6$, this gives an estimate for the circuit depth required to implement a single trial of the symmetric-key generation scheme.

The results of these tests show that the LAQW model presents significantly reduced circuit depths over values of $t \in \{8, \dots, 20\}$. The average reduction $\bar{r}_{\text{Generic}}$ in circuit depth when transpiling the circuits for \texttt{GenericBackendV2} was approximately $\bar{r}_{\text{Generic}}\approx88.28\%$, with a standard deviation of $\sigma_{\text{Generic}}=0.39\%$. When transpiling the circuits for IBM's \texttt{Torino} QPU, the average reduction $\bar{r}_{\text{Torino}}$ was approximately $\bar{r}_{\text{Torino}}\approx 87.81\%$ with a standard deviation of $\sigma_{\text{Torino}}=1.33\%$. Since the relative circuit depth reductions are close to each other for both backends, this indicates that the relative depth reduction magnitude is not substantially affected by considering a different qubit connectivity.

\begin{figure}[t]
\centering
\includegraphics[width=0.9\linewidth]{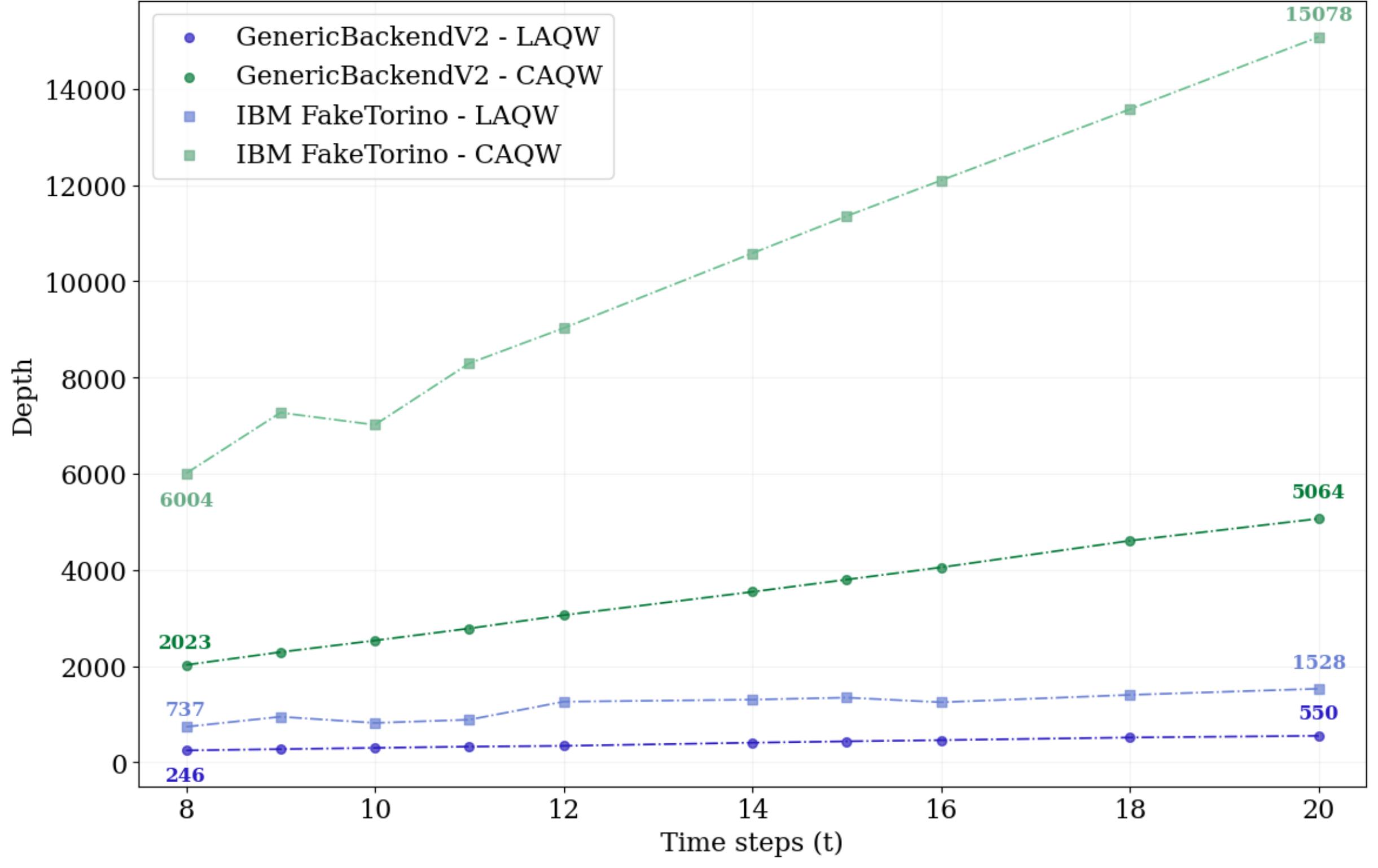}
\caption{Average circuit depth of the LAQW and CAQW models over time steps $t \in \{8,\dots,20\}$. The circuits were constructed using Qiskit and transpiled for \texttt{GenericBackendV2} and \texttt{FakeTorino} backends. Plots with cycle markers represent the LAQW (blue) and CAQW (green) circuit depth using the \texttt{GenericBackendV2} simulator, while the plots with square markers represent the circuit depth using the \texttt{FakeTorino} backend simulator.}
\label{fig:circuitComparisons}
\end{figure}

The circuit depths for different number of time steps $t$ can be seen in Figure \ref{fig:circuitComparisons}. The figure shows a substantial reduction in the circuit depth required by the LAQW. When using the noiseless \texttt{GenericBackendV2} simulator, the LAQW circuit exhibits a maximum depth of $550$ for $t=20$ time steps, compared to 5 064 for the CAQW. This efficiency gap widens under noisy simulation, where the compiler introduces additional \texttt{SWAP} gates to enable quantum operations between non-adjacent qubits on the hardware. 

The gap is to be expected when the circuit implementations of the LAQW and CAQW are considered. As described in Section \ref{sec3.2}, the CAQW's shift operator is implemented using controlled decrement and increment circuits, which require numerous interactions between the coin qubit and specific pairs of position qubits. On hardware with limited qubit connectivity, non-local interactions result in a large number of \texttt{SWAP} gates that increase both circuit depth and gate count. Conversely, the LAQW's QFT-based shift operator performs a global transformation acting on the entire position register. Although the QFT and QADD$\pm$ operations still require non-local interactions, the QADD$\pm$ operations have fewer control qubits than the corresponding increment and decrement operations. Therefore, the compiler is able to optimize the qubit interactions with fewer \texttt{SWAP} gates.

This reduction of approximately $88\%$ in circuit depth highlights the LAQW as a more practical candidate for implementing QW-based protocols on near-term quantum hardware. However, it is important to note that the circuit depths of the LAQW method are still too high for the capabilities of current hardware, as characterized by~\cite{proctor2024benchmarkingquantumcomputers}. 

\subsection{Parameter sensitivity comparison}

In order to use quantum walks as chaotic systems, the probability distributions that the quantum walk models produce have to be sensitive to the initial values given for the model. This means that even a small change in the initial values should result in the quantum walk model producing a significantly different probability distribution.  

We use the \textit{Hellinger distance}, defined as the probabilistic counterpart of the standard Euclidean distance, as a metric for measuring how different two probability distributions in a discrete probability space are from each other. The Hellinger distance of two probability distributions of $k$ amplitudes $P=(p_1,...,p_k)$ and $Q=(q_1,...,q_k)$ is given by 
\begin{equation}
    H(P,Q)=\frac{1}{\sqrt{2}}\sqrt{\sum_{i=1}^k(\sqrt{p_i}-\sqrt{q_i})^2}. \label{eq:HellingerDist}
\end{equation}

The tests were conducted by randomly picking 1 000 sets of initial values from a uniform distribution. For each of these sets of initial values, the unitary evolution operator $U$ of Eq. \eqref{CAQWUnitary} was computed in matrix form using NumPy~\cite{harris2020array} for both the LAQW and CAQW models. The walk was then simulated using NumPy by applying $U$ to the initial state $\ket{\psi_0}$ in vector form $t$ times. Each position $\ket{x,y}$ was then assigned a measurement probability according to \begin{equation}
    p_{x,y}=|\braket{x,y|U^t|\psi_0}|^2.
\end{equation} 
This way, the probability distributions produced by the LAQW and CAQW models can be numerically obtained. After this, one or multiple initial values were varied by one percent of the original value (time steps $t$ were varied by $\pm1$ steps) and the new varied probability distributions were computed in a similar way to the original probability distribution. After obtaining the original and the varied probability distribution, the Hellinger distance between them was calculated by Eq. \eqref{eq:HellingerDist}. The average Hellinger distances for the LAQW and CAQW models with different parameter changes can be seen in the first 5 rows of Table \ref{tab:Parameter sensitivity}. Additionally, the average distance between probability distributions corresponding to completely unrelated sets of initial parameters was computed. This was done by randomly picking 2 000 sets of initial values from a uniform distribution and numerically computing the probability distributions that the models produce as described previously. After this, the probability distributions were divided into pairs, and the Hellinger distances were calculated between each pair of distributions. The average Hellinger distances for both models can be seen in the last row of Table \ref{tab:Parameter sensitivity}.

\begin{table}[h]
\centering
\begin{tabular}{|c|c|c|c|}
\hline
\textbf{Parameter change} & \begin{tabular}{c}
    \textbf{LAQW Hellinger} \\
     \textbf{distance average $H_L$}
\end{tabular} & \begin{tabular}{c}
     \textbf{CAQW Hellinger} \\
     \textbf{distance average $H_C$}
\end{tabular} & $\frac{H_L}{H_C}$ \\ \hline
$\theta_0\pm 1\%$  & 0.05599 & 0.04278 &1.30120\\ \hline
$\theta_1\pm 1\%$  & 0.05685 & 0.04354 & 1.29630\\ \hline
$\alpha\pm 1\%$  & 0.00978 & 0.00814 & 1.18976\\ \hline
$t\pm 1$  & 0.40044 & 0.55526 & 0.71974 \\ \hline
$\theta_0,\theta_1,\alpha\pm 1\%$  & 0.09875 & 0.07723 & 1.28724 \\ \hline
 \begin{tabular}{c}
     \text{Different initial} \\
     \text{parameters}
\end{tabular} & 0.50257 & 0.58091 & 0.85375 \\ \hline

\end{tabular}
\caption{Average Hellinger distances $H_L$ and $H_C$ of the probability distributions produced by LAQW and CAQW models when initial parameters are slightly varied (first 5 rows) and when two sets of different initial parameters are used (last row).}
\label{tab:Parameter sensitivity}
\end{table}

It can be observed that the LAQW model is more sensitive when it comes to small changes in the $\theta_0$, $\theta_1$ and $\alpha$ parameters, with the average Hellinger distance being higher by $30.12\%$, $29.63\%$, and $18.98\%$, respectively. When all three of these parameters are slightly changed, the LAQW has on average $28.72\%$ higher Hellinger distances than the CAQW. However, the CAQW model is more sensitive to changes in the number of time steps. This can be seen from the average Hellinger distance of the LAQW being $20.03\%$ lower than the average Hellinger distance of the CAQW when $t$ is changed by one. This can be explained by the LAQW walker having the option to stay in place, which means that the walk spreads more slowly on the graph when a step is taken. 

Additionally, for two completely random sets of initial values, the LAQW model has a $14.63\%$ smaller average Hellinger distance value than the CAQW model. These results show that, even though the LAQW model has greater sensitivity for some intial parameters, with different sets of initial values, the CAQW has a higher average Hellinger distance. This means that, on average, the probability distributions produced by the CAQW are more random. Therefore, a trade-off can be identified between the circuit depth of the chosen model and the randomness of the results produced by the chosen model.

\subsection{Statistical tests conducted}
\label{statistical_analysis}

The main objective of this section is to assess the statistical quality of our LAQW model's generated raw and key bitstrings. We are also interested in comparing these results with the statistical evaluation of the CAQW model to learn more about the statistical behavior and entropy characteristics of different quantum walk models. We conduct a series of independent trials $\mathcal{T}$ and evaluate both the raw and extracted key bitstrings using well-established randomness tests to check that they exhibit the unpredictability and lack of structure that is expected from a physical entropy source. 

It should be noted that, in our key generation setup, we fixed the final key length to $128$ bits, as this is a common choice for many symmetric-key algorithms. For instance, 128-bit keys are used in many encryption schemes, such as AES-128, since that length is sufficient for many secure communication protocols while remaining more computationally efficient~\cite{1250456}. However, longer keys (i.e., 256 bits or more) used in higher-security settings could also be chosen. 

Following the bitstring generation protocol detailed in Section \ref{bitstring_generation}, the LAQW produced a raw bitstring of length $B_{\mathrm{raw}} =5120$ bits for $r=10$ runs. We chose to evaluate the performance of a bitstring composed of multiple per-run bitstrings to observe if the statistical randomness of the combined bitstring is of high quality. If so, a raw bitstring can be large in length, making it possible to also generate larger key bitstrings. To evaluate our results, we generated a raw bitstring $\mathcal{T}=100$ times. For each $\mathcal{T}_i$, we estimated the conservative min-entropy, which yielded an average value of $h_{\min}=2.781$ bits per block. This resulted in an average of 1 780 extractable bits. Compared to the LAQW results, the CAQW model exhibited an average min-entropy of $h_{\min}=1.947$ bits per block. With a raw bitstring length of $3920$ bits for $r=10$ runs, this corresponds to 954 extractable bits. Although the length is enough for the extraction of a 128-bit key, the available min-entropy per block using the CAQW model is noticeably lower. This difference can be attributed to the underlying dynamics of the two quantum walk models. As a result, both the min-entropy $h_{\min}$ and the number of extractable bits are less for the CAQW case.

Although the min-entropy analysis allows for a certain number of bits to be extracted to form a key, this does not ensure that the generated bitstrings are free from statistical patterns or correlations. For this reason, we incorporated a subset of the NIST Statistical Test Suite~\cite{8966} that consists of 15 tests, all of which are designed to evaluate different aspects of the quality of true and pseudorandom number generators. For instance, these tests check for global and local biases in the proportion of ones, unusually long runs of zeros and ones, periodic or structured components in the frequency domain, and cumulative deviations. In our work, we applied the frequency (monobit) test, block frequency test, runs test, longest run of ones test, spectral test, and cumulative sums test to both the raw bitstring and the extracted $128$-bit key. By combining a subset of these complementary tests, we have been able to gather a more thorough assessment of whether our generated bitstrings are statistically consistent with an ideal random source. In the following, we give a brief description of each test used and provide the statistical results in Table \ref{table:1}.

\paragraph{NIST SP 800-22}
To interpret the results of the NIST tests, we adopted the standard value $\alpha = 0.01$ that defines the \textit{level of significance}, or the probability of incorrectly rejecting the null hypothesis that a bitstring sequence is random. This value is not to be mistaken for the initial parameter $\alpha$, which determines the initial state of the coin qubits. Each test produces a probability value $p$ that determines how strongly the results align with this null hypothesis. Specifically, probability values $p< 0.01$ suggest that the bitstring fails the test and is a non-random sequence, whereas $p\ge 0.01$ suggests the sequence is random with $99\%$ confidence. Using this criterion, we evaluated the performance of the generated bitstrings using the subset of NIST tests introduced above. We briefly describe the tests and report the p-values obtained for both the raw and key bitstrings. More details on how each test is implemented can be found in~\cite{8966}.

\begin{enumerate}
    \item The \textit{Frequency (monobit) test} measures the global balance between zeros and ones across the bitstring sequence. For a truly random sequence, the fraction of ones should be approximately equal to the number of zeros. The p-value is based on the complementary error function of $s_{\text{obs}}$, which is the normalized sum of the bits with $0$ mapped to $-1$ and $1$ mapped to $1$. A value of $s_{\text{obs}}$ close to zero means that there is a more balanced number of $0$ and $1$ bits. If $p \ge 0.01$, the bitstring sequence is considered to be random. We obtained $p = 0.118$ for the raw bitstring and $p=0.052$ for the key bitstring, indicating no significant bias in the bit values in either case.
    
    \item The \textit{Block Frequency test} is based on the Frequency test and examines the local behavior in the bitstring sequence. Specifically, it assesses whether the frequency of ones within a fixed $M$-length block of the bitstring is close to $M/2$. This is a way to detect biases in the bitstring that might not be noticed globally. The test statistic $\chi^2$ sums together the square of the deviations of the number of $1$s in each block from $1/2$. Larger values of $\chi^2$ indicate stronger local bias. The p-value for this test is based on the incomplete gamma function as defined in~\cite{8966}. With $\chi^2 = 269.600, p = 0.268$ for the raw bitstring and $\chi^2 = 5.600, p = 0.469$ for the key bitstring, both bitstring sequences have deviations that are consistent with the statistical fluctuations expected from a random sequence.
    
    \item The \textit{Runs test} examines the number of \textit{runs} across the entire sequence, where a run is defined as the maximal sequence of consecutive identical zeros or ones. The value $V_n$ denotes the total number of such runs. For a random bitstring sequence with a certain proportion of ones, an expected number of runs can be estimated. If $V_n$ is higher or lower than expected, it could indicate non-random oscillatory behavior, such as excessive (or too little) clustering. The p-value is based on the complementary error function of $V_n$. In our results, the value of $V_n$ for the raw bitstring is $V_n = 2454$ with $p = 0.003$. The large value of $V_n$ suggests that oscillations in the bit values occur more often than expected for a random sequence at the chosen significance level. In contrast, the key bitstring has $V_n = 58$ with $p = 0.454$, which is consistent with the expected number of runs for a random sequence.
    
    \item The \textit{Longest Run of Ones test} evaluates the distribution of the longest run of consecutive ones within $M$-length blocks of a sequence. For each block, $V_n$ is recorded, and the distribution over all blocks is then compared to the $\chi^2$ distribution expected for a random sequence. The value $\chi^2$ is the measure of the discrepancy between the observed and expected distribution of run lengths, where larger values of $\chi^2$ suggests excessive clustering of ones. The p-value is determined by the incomplete gamma function. For the raw bitstring, we obtained the values $\chi^2 = 6.087, p = 0.107$ with a maximum observed run length $V_{\max}$ of $8$ ones. For the key bitstring, $\chi^2 = 10.620$, with $p = 0.014$ with a maximum run $V_{\max}$ of $5$ ones. While the $\chi^2$ values of both bitstring sequences are relatively small, the p-value for the key bitstring is closer to the rejection threshold.
    
    \item The \textit{Spectral test} can detect periodic or repetitive patterns in a sequence that deviate from randomness. First, bits $0$ and $1$ are mapped to $-1$ and $1$, respectively, after which a discrete Fourier transform is applied to transform the sequence into the frequency domain. The number of peaks observed in the frequency spectrum that exceed a given threshold corresponding to $(1-\alpha)$ are compared with the number of peaks expected from a random sequence. The value $d$ is then the normalized difference between the observed and expected number of peaks. If values of $d$ are far from zero, this could suggest periodic structure within the bitstring sequence. The p-value in this test is based on the complementary error function of $d$. For the raw bistring, we obtained $d=-0.910$ with $p = 0.362$, indicating no significant evidence for periodicity. For the key bitstring, $d=0.090$ with $p=0.931$, which is also indicative of randomness. However, in this case the results are not statistically valid, since the standard sequence length for this test is $> 1000$ bits. 
    
    \item The \textit{Cumulative Sums test} evaluates whether partial sums of a sequence exhibits systematic drift. Specifically, bits in the bitstring sequence are first mapped from $0$ $(1)$ to $-1$ $(1)$. Then, the cumulative sum is computed along the bitstring and the maximum absolute deviation from zero is recorded. The values $z_f$ and $z_b$ measure this maximum deviation when moving in forward and backward directions along the bitstring sequence, respectively. Large $z$ values indicate an accumulation of an imbalance of zeros or ones, such as there being too many ones at the beginning of the sequence. The p-value for this test comes from the Standard Normal Cumulative Probability Distribution Function defined in~\cite{8966}. For the raw bitstring, we found $z_f=1.789$ and $z_b=1.565$ with a p-value of $p=1.0$. Similarly for the key bitstring, $z_f=2.210$ and $z_b=1.945$ with $p=1.0$. These results suggest that both bitstring sequences show no significant cumulative biases in either direction.
\end{enumerate}

\begin{table}[h!]
\centering
\begin{tabular}{|c|c|c|c|c|}
    \hline
    \multicolumn{5}{|c|}{\textbf{NIST SP 800-22}} \\
    \hline
     \textbf{Test} & \textbf{Raw bitstring} & $p \geq 0.01$ & \textbf{Key bitstring} & $p \geq 0.01$\\ 
     \hline
     Frequency  & \begin{tabular}{@{}c@{}}$s_{\mathrm{obs}}=1.565$ \\ $p=0.118$\end{tabular} & pass & \begin{tabular}{@{}c@{}}$s_{\mathrm{obs}}=1.945$ \\ $p=0.052$\end{tabular} & pass \\
     \hline
     Block frequency &  \begin{tabular}{@{}c@{}}$\chi^2=269.6$ \\   $p=0.268$\end{tabular}  & pass & \begin{tabular}{@{}c@{}}$\chi^2=5.6$ \\ $p=0.469$\end{tabular} & pass \\
     \hline
     Runs & \begin{tabular}{@{}c@{}}$V_n=2454$ \\ $p=0.003$\end{tabular} & fail & \begin{tabular}{@{}c@{}}$V_n=58$ \\ $p=0.454$\end{tabular} & pass\\
    \hline
    Longest run of ones & \begin{tabular}{@{}c@{}}$\chi^2=6.087$ \\ $p=0.107$\end{tabular} & pass &  \begin{tabular}{@{}c@{}}$\chi^2=10.620$ \\ $p=0.014$\end{tabular} & pass\\
     \hline
     Spectral (DFT) & \begin{tabular}{@{}c@{}}$d=-0.91$ \\ $p=0.362$\end{tabular}  & pass & \begin{tabular}{@{}c@{}}$d=0.09$ \\ $p=0.931$\end{tabular} & $\mathrm{pass}^\dagger$\\
     \hline
     Cumulative sums & \begin{tabular}{@{}c@{}}$z_f=1.789$ \\ $z_b=1.565$ \\ $p=1.0$\end{tabular}  & pass  & \begin{tabular}{@{}c@{}}$z_f=2.210$ \\ $z_b=1.945$ \\ $p=1.0$\end{tabular} & pass \\
     \hline
\end{tabular}
\caption{Summary of selected NIST statistical tests~\cite{8966} for both the raw and key bitstrings. A test is marked ``pass'' when the p-value is greater than the significance level, $\alpha=0.01$. The $\dagger$ symbol denotes that, although the key bitstring passes the Spectral test, it is not sufficiently long for the results to be considered statistically valid.}
\label{table:1}
\end{table}

The statistical analysis indicates primarily that the generated bitstrings successfully passed most of the applied randomness tests at both the raw and post-processing (rounding and prime-modulus mapping) stages of the key generation process. These results provide evidence that the LAQW, in combination with the post-processing scheme, can effectively produce a bitstring sequence whose behavior is consistent with randomness. While there were some minor deviations observed in specific tests, such as the Runs test for the raw bitstring, these deviations were mitigated during the entropy estimation and bit extraction stages described in Section \ref{min_entropy}. The resulting key bitstring exhibited no statistically significant deviations from randomness for the tests we applied.

\begin{table}[h!]
\centering
\begin{tabular}{|c|c|c|c|c|}
    \hline
    \multicolumn{5}{|c|}{\textbf{NIST SP 800-22}} \\
    \hline
    \textbf{Test} & \textbf{LAQW} & $p \geq 0.01$ & \textbf{CAQW} & $p \geq 0.01$\\ 
    \hline
    Frequency  & \begin{tabular}{@{}c@{}}$s_{\mathrm{obs}}=1.565$ \\ $p=0.118$\end{tabular} & pass & \begin{tabular}{@{}c@{}}$s_{\mathrm{obs}}=0.256$ \\ $p=0.798$\end{tabular} & pass \\
    \hline
    Block frequency &  \begin{tabular}{@{}c@{}}$\chi^2=269.6$ \\ $p=0.268$\end{tabular}  & pass & \begin{tabular}{@{}c@{}}$\chi^2=252.8$ \\ $p=0.004$\end{tabular} & fail \\
    \hline
    Runs & \begin{tabular}{@{}c@{}}$V_n=2454$ \\ $p=0.003$\end{tabular} & fail &  \begin{tabular}{@{}c@{}}$V_n=1906$ \\ $p=0.085$\end{tabular} & pass\\
    \hline
    Longest run of ones & \begin{tabular}{@{}c@{}}$\chi^2=6.087$ \\ $p=0.107$\end{tabular} & pass &  \begin{tabular}{@{}c@{}}$\chi^2=1.613$ \\ $p=0.656$\end{tabular} & pass\\
    \hline
    Spectral (DFT) & \begin{tabular}{@{}c@{}}$d=-0.91$ \\ $p=0.362$\end{tabular}  & pass & \begin{tabular}{@{}c@{}}$d=-0.42$ \\ $p=0.675$\end{tabular} & pass\\
    \hline
    Cumulative sums & \begin{tabular}{@{}c@{}}$z_f=1.789$ \\ $z_b=1.565$ \\ $p=1.0$\end{tabular}  & pass  & \begin{tabular}{@{}c@{}}$z_f=0.607$ \\ $z_b=0.591$ \\ $p=1.0$\end{tabular} & pass \\
    \hline
\end{tabular}
\caption{Summary of selected NIST statistical tests comparing the LAQW and CAQW models. A test is marked ``pass'' when the p-value is greater than the significance level $\alpha=0.01$.}
\label{table:2}
\end{table} 

\paragraph{Comparison with CAQW}
We applied the same subset of NIST SP 800-22 tests to the raw bitstrings generated by the CAQW model. As summarized in Table \ref{table:2}, the CAQW raw bitstring passed five out of the six tests, failing only the block frequency test with $p = 0.004$. This failure indicates the presence of local imbalances in the proportion of zeros and ones within fixed-length blocks. In comparison, the LAQW model fails only the Runs test with $p = 0.003$, which suggests a tendency toward more frequent alternations between bit values than expected for a random sequence. However, both models have p-values well above the $p \geq 0.01$ in the Frequency, Longest Run of Ones, Spectral, and Cumulative Sums tests, indicating no strong evidence of clustering, periodicity, or cumulative drift in either case.

While both LAQW and CAQW models are able to produce raw bitstrings with strong overall randomness properties, the specific test failures may provide insight into the differences in their statistical behavior. However, analyzing this statistical behavior on a deeper level would require further research and is currently beyond the scope of this paper.

\subsection{Reproducibility}
\label{reproducibility}
A fundamental requirement for any deterministic symmetric-key generation scheme is \textit{reproducibility}. This ensures that a sender and receiver can establish a shared secret without having to communicate anything more than the initial parameters used in the quantum walk. To validate this property for our scheme, we executed $\mathcal{T}=100$ independent \textit{trials} of the full key generation process. For each trial, we generated two raw bitstrings $B_{\mathrm{raw},i}$ and $B_{\mathrm{raw},j}$ using the same initial parameters and extracted a 128-bit key from each by following the procedure discussed in Section \ref{bitstring_generation}.

\begin{figure}[h]
    \centering
    \includegraphics[width=0.9\linewidth]{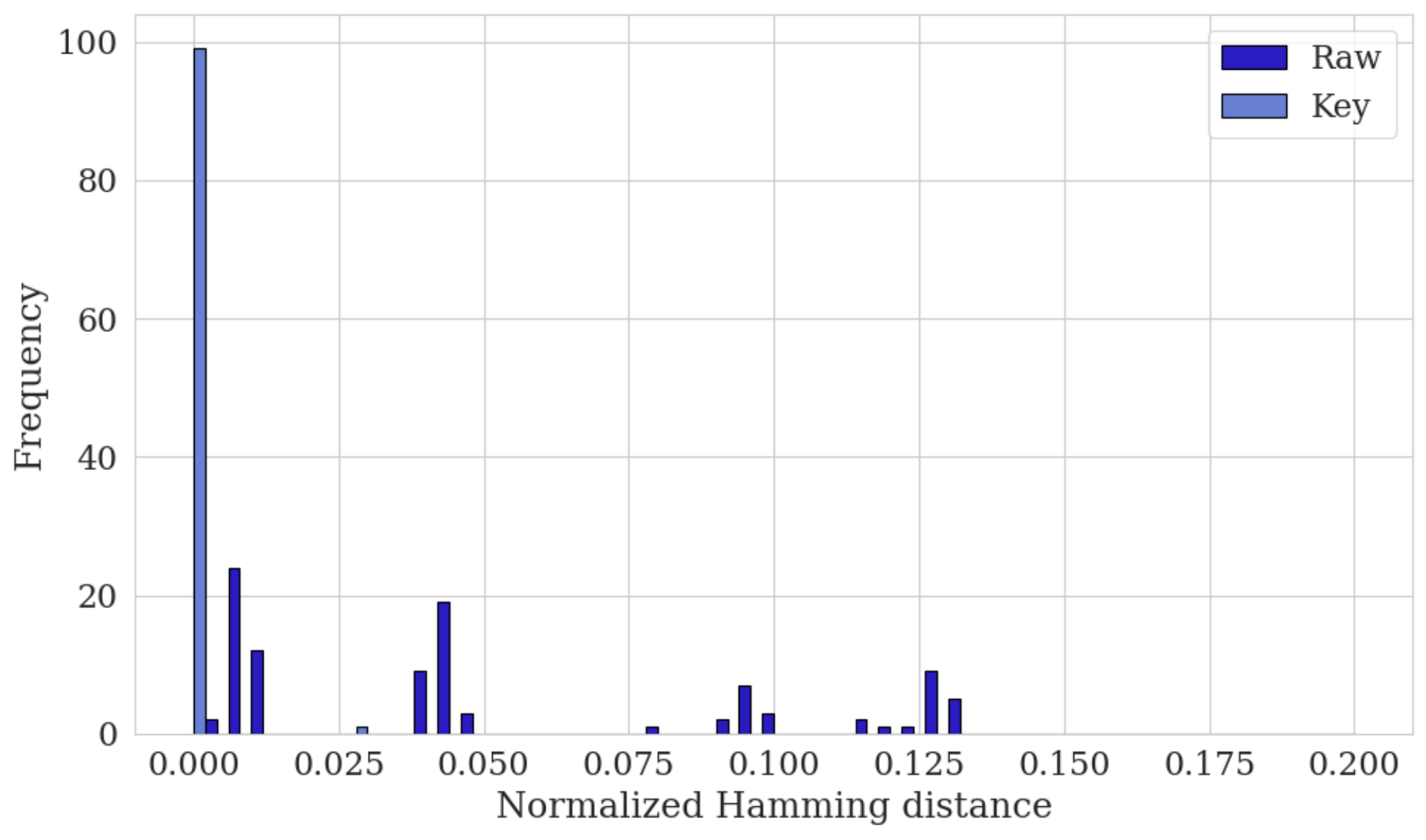}
    \caption{Frequency of the normalized Hamming distances for the raw (dark blue) and key (light blue) bitstrings of the LAQW model across $\mathcal{T}=100$ trials. Each trial consists of two bitstrings being independently generated and compared. The bars representing each model are plotted side-by-side rather than overlapping.}
    \label{fig:reproducibility}
\end{figure}

Next, we quantified the similarity between the bitstrings from $\mathcal{T}_i$ by using the \textit{Hamming distance}~\cite{hamming}: 
\begin{equation}
    H\,(B_i,B_j) = \sum_{k=1}^{L}B_i(k)\oplus B_j(k),
\end{equation}
where $B_i$ is defined as $B_i = (B_i(1),B_i(2),\dots,B_i(L))\in\{0,1\}^L$ and $B_j$ is defined the same way. As shown in Figure \ref{fig:reproducibility}, the normalized Hamming distance across all pairs of raw bitstrings over 100 trials is 
$0.0490$, where $L_{\text{raw}}=5120$ bits. We used the normalized Hamming distance to better visualize the results. The larger Hamming distances of the raw bitstrings are primarily caused by the counts $c_{r,i}$ being near the rounding boundaries, and therefore being mapped to different integer bins in different trials. This leads to differing byte values after the prime-modulus mapping.

The final $128$-bit keys exhibited a normalized Hamming distance of $0.0003$ across 100 independent trials. Although there was one trial where the normalized Hamming distance of the key was non-zero (with a Hamming distance of 4 bits), a simple \textit{majority-vote rule} could be implemented to ensure identical keys for both communicating parties with high probability. Specifically, if multiple keys are generated from the same initial parameters, the most frequently-occurring key over multiple trials would be selected. However, the results of implementing this ``majority-vote'' key will not be discussed further, as we did not include this step in our setup.

\begin{figure}[h]
\centering
\includegraphics[width=0.9\linewidth]{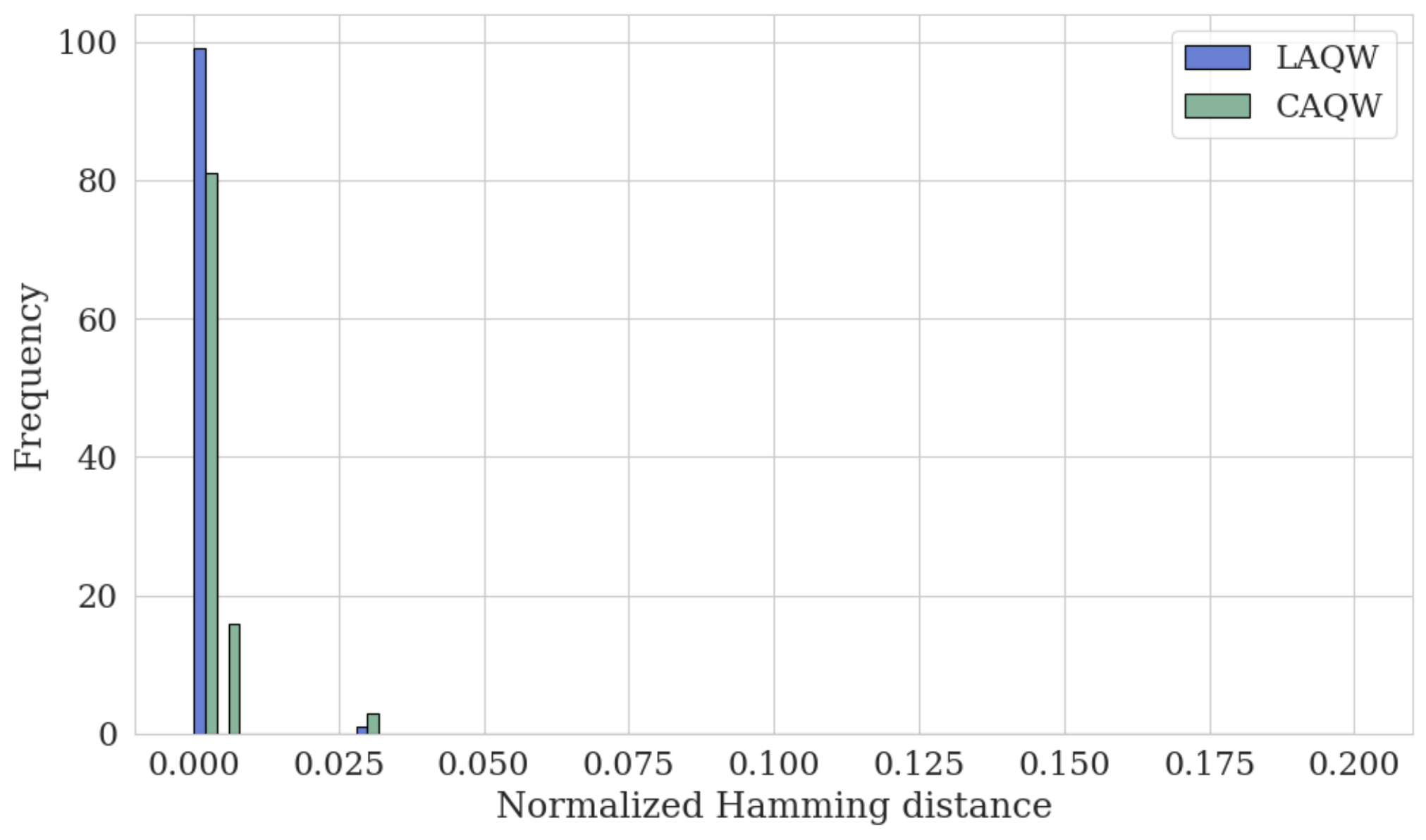}
\caption{Frequency of the normalized Hamming distances for the key bitstrings between the LAQW and CAQW models across $\mathcal{T}=100$ trials. The bars representing each model are plotted side-by-side rather than overlapping.}
\label{fig:LAQWvsCAQW}
\end{figure}

\paragraph{Comparison with CAQW}
Following the same process as with the LAQW model, we performed $\mathcal{T}=100$ independent trials of the key generation scheme using the CAQW model. The average CAQW normalized Hamming distance for the raw and key bitstrings was found to be $0.0210$ and $0.0022$, respectively. Figure \ref{fig:LAQWvsCAQW} shows the comparisons between both models. So, despite both models having raw bitstrings with non-zero Hamming distances, the key extraction process was able to successfully suppress these differences. This suggests that both quantum walk models are able to reliably reproduce identical keys using our key generation scheme when starting with the same initial parameters.

\subsection{Effects of Noise}
\label{effects_noise}

In the context of symmetric-key generation, noise can influence the reproducibility of the generated bitstrings by altering the probability distributions of the walk, making it more difficult to get the same distribution even when using identical initial parameters. In quantum computers, noise is encountered primarily in the form of gate infidelities, decoherence, qubit cross-talk, and readout errors. We evaluated the resilience of our proposed key generation scheme to realistic noise using IBM's \texttt{FakeTorino} backend~\cite{IBMQuantum} following the steps from the previous section. We executed $\mathcal{T}=100$ independent trials using the \texttt{FakeTorino} backend and quantified the similarity of two bitstrings from trials $\mathcal{T}_i$ using the Hamming distance, as described previously. Lastly, we compared these results for the raw and key bitstrings when using the \texttt{FakeTorino} backend with the results found in Section \ref{reproducibility} using the \texttt{GenericBackendV2} simulator.

Across $\mathcal{T}=100$ independent trials using the \texttt{FakeTorino} backend, our LAQW scheme resulted in an average of 1 644 extractable bits, which is the same number of bits as when using the noiseless \texttt{GenericBackendV2} simulator. For the raw bitstrings, the introduction of noise led to a slight increase in the average normalized Hamming distance, from $0.0490$ with \texttt{GenericBackendV2} to $0.0500$ with \texttt{FakeTorino}. In contrast, the key bitstrings exhibited extremely low Hamming distance in both cases. For the \texttt{FakeTorino} backend, the average normalized Hamming distance was found to be $0.0002$, compared to $0.0003$ for the noiseless simulator. Across all trials with the noisy backend, only one single instance of a non-zero Hamming distance of 2 bits was observed for the key bitstrings. Based on these results, the proposed key generation scheme remains intact under realistic device noise.

\section{Conclusions}
\label{sec6}

In this paper, we proposed a LAQW model that can be used for creating chaotic probability distributions that are highly sensitive to the initial values of the walk. We focused on the applicability of this model as a subroutine for different symmetric obfuscation schemes in place of the previous quantum walk models used in~\cite{CAQWmed,AQWPRNG2020,li2017controlledalternatequantumwalks}. The main improvement in our proposed model is that the quantum circuit implementation of the LAQW model requires a circuit depth which scales as $\mathcal{O}(n^2+nt)$, whereas the CAQW circuit implementation requires a circuit depth that scales as $\mathcal{O}(n^2t)$. This was discussed in detail in Section \ref{sec2}. On average, the LAQW model provided a circuit depth reduction of $\approx 88\%$ when compared to the CAQW model.

The LAQW had $30.12\%$, $29.63\%$, and $18.98\%$ higher average Hellinger distances between perturbed and non-perturbed probability distributions than the CAQW when the parameters $\theta_1$, $\theta_2$, and $\alpha$ were changed by $\pm 1\%$, respectively. However, the LAQW had $20.03\%$ lower average Hellinger distance between perturbed and non-perturbed probability distributions than the CAQW when the amount of time steps was changed by $\pm 1$. From these results, we learn that the LAQW is more sensitive to small changes in parameters that affect the movement of the walk, while the CAQW is more sensitive in the number of time steps the walker takes. Nevertheless, due to the significantly shallower circuit depths required to implement the LAQW model, the LAQW model represents a more promising algorithm to be used instead of the CAQW for encryption applications such as in~\cite{CAQWmed,AQWPRNG2020,li2017controlledalternatequantumwalks}.

As described in Section \ref{sec2.2}, the AQW and CAQW models have been used in previous literature to generate chaotic maps for symmetric image obfuscation schemes. In this paper, we have provided an example of how the approximate probability distribution generated by a quantum walk on a graph can be used to create a reproducible cryptographic key. In general, the probability distribution can be measured on a quantum computer, for example, by sampling single observations until an approximation of the probability distribution is obtained from the set of measurements. However, this gives only an approximation of the underlying probability distribution. As the number of executions of the circuit increases, so does the accuracy of the approximation. In the future, the optimal approach for generating a cryptographic key with the approximated probability distribution should be studied more, as it is a critical part of how the LAQW, AQW, and CAQW models can be used on quantum hardware for symmetric obfuscation purposes. 

It is also important to comment on the trade-offs we considered during the construction of our proposed symmetric-key generation scheme. First, there is a trade-off between the accuracy of the obtained probability distribution and the number of times the circuit is executed. Additionally, if the size of the two-dimensional lattice is increased, the number of circuit executions also has to be increased to maintain the accuracy of the probability distribution. Similarly, since the scheme uses the approximated probability distribution of the quantum walk, one must decide how to balance between the reproducibility of the key using identical initial parameters and the sensitivity of the key with respect to the initial parameters. These trade-offs should also be further researched in order to better understand the usefulness of the quantum walk models in chaos-based cryptographic applications, such as image obfuscation. 

\paragraph{Acknowledgements} N.G. received funding from the Quantum Doctoral Programme. All the authors acknowledge funding by Business Finland for the project 100/31/2024 BLimPQC, and the use of IBM Quantum services~\cite{IBMQuantum} for this work.

\bibliographystyle{unsrtnat}
\bibliography{laqw_bibliography}

\end{document}